\documentclass[twocolumn,preprintnumbers,amsmath,amssymb,showpacs]{revtex4}
\usepackage{graphicx,caption,subcaption,float}
\begin{document}

\title{Crucial events, randomness and multi-fractality in heartbeats }

\author{ Gyanendra Bohara $^{1}$, David Lambert $^{1}$, Bruce J. West $^{2}$, Paolo Grigolini $^{1}$}
\address{$^1$ Center for Nonlinear Science, University of North Texas, P.O. Box 311427,
	Denton, Texas 76203-1427, USA\\ }
\address{$^2$ Information Science Directorate,  Army Research Office, Research Triangle Park,  NC 27708}

\begin{abstract}
We study the connection between multi-fractality and crucial events. Multi-fractality is frequently used as a 
measure of physiological variability.  Crucial events are known to play a fundamental role in the transport of information between complex networks. To establish a connection we focus on the special case of heartbeat time series and on the search for a diagnostic prescription to distinguish healthy from pathologic subjects. Over the last twenty years two apparently different diagnostic techniques have been established: the first is based on the observation that the multi-fractal spectrum of healthy patients is broader than the multi-fractal spectrum of pathologic subjects;  the second 
is based on the observation that heartbeat dynamics are a superposition of crucial and Poisson events, with pathologic
patients hosting Poisson events with larger probability than the healthy patients.  In this paper, we prove that increasing the percentage of Poisson events hosted by heartbeats has the effect of making their multi-fractal spectrum narrower, thereby establishing that the two different diagnostic techniques are compatible with one another and, at the same time, establishing a dynamic interpretation of multi-fractal processes that has been previously overlooked.\end{abstract}


\maketitle

\section{Introduction}

The hypothesis that multi-fractality is a significant property of physiological processes gained attention in the literature
of the last 20 years.  Ivanov \emph{et al} \cite{stanley} initiated this interest using wavelets to analyze the heartbeat data of several patients, some healthy and some affected by congestive  heart failure.  They determined that the main difference between the healthy and non-healthy is that the healthy subjects have a significantly broader multi-fractal spectrum. The multi-fractal approach  \cite{stanley} is an efficient way to measure  cardiovascular \emph{variability} \cite{variability1}, referred to as heart rate variability (HRV),  the proper  treatment of which is still the object of intense discussions \cite{variability2}. 

The statistical analysis of heartbeat sequences, as well as that of other
physiological processes, is carried out by properly processing suitable 
time series. Each time series corresponds to a single individual who is unique, thereby raising the challenging problem of determining how to establish a connection with the Gibbs ensemble perspective, which requires  averages to be taken over identical copies of the same system.   This dilemma is settled by assuming that different portions of the single time series can be interpreted as identical copies of the same process, corresponding to different initial conditions. A well known analysis technique of this kind is  \emph{Detrended Fluctuation Analysis} (DFA), \cite{DFA1,DFA2}. 
Due, in part,  to the growing interest in multi-fractality  \cite{vulpiani}, Kantelhardt \emph{et al.} \cite{stanley3}
extended DFA so as to make it possible to extract from it 
multi-fractal information, through the spectral density $f(\alpha)$ which 
often has the form of a broad inverted parabola that is expected to become very narrow and centered on the scaling index $\alpha = 0.5$ in the ordinary Poisson case. We refer to the algorithm developed in \cite{stanley3} as Multi Fractal Detrended Fluctuation Analysis (MFDFA). MFDFA is adopted to discuss the transmission of multi-fractality from a complex stimulus to another complex network \cite{deligniers}, both being characterized by a broad
$f(\alpha)$ spectrum. 

The main purpose of this article is to uncover the dynamical origin of a broad $f(\alpha)$ spectrum by moving from the specific case of HRV to the general properties of non-Poisson time series. To achieve this, we follow  the search  for a diagnostic distinction between healthy and pathological subjects. The goal, however,  is to obtain  a better understanding of the dynamical origin of multi-fractal variability. Significant insights about this dynamic origin would attract  general interest to the improvement of diagnostic techniques.  One possible road to the solution of this problem can be found by noticing that in 2002  Allegrini \emph{et al} 
 \cite{allegrini}  used the detection of crucial events as the main criterion to distinguish healthy (with a broad $f(\alpha)$ spectrum) from unhealthy (with narrow $f(\alpha)$ spectrum) patients. For a proper definition of crucial events we adopt the theoretical perspective established in earlier work, see for example \cite{brain}, defining  the  \emph{crucial events} on the basis of the time interval between the occurrence of two consecutive events. The time interval between two consecutive events is described by a waiting time probability density function (PDF) 
 $\psi(\tau)$. In the case of crucial events $\psi(\tau)$ has the asymptotic inverse power law (IPL) structure:
\begin{equation} \label{crucial1}
\psi(\tau) \propto \frac{1}{\tau^{\mu}}
\end{equation} 
with $\mu < 3$. 
The time intervals between two different pairs of
consecutive events are not correlated:
\begin{equation} \label{crucial2}
\left<\tau_i \tau_j\right> \propto \delta_{ij}. 
\end{equation}
The occurrence of crucial events plays an important role in the transport of information from one to another complex network \cite{west}


  It is important to discuss the dynamical origin of events of this kind. Crucial events are a manifestation of cooperative interactions between the units of a complex network that is expected to lead to a spontaneous organization process, usually called Self Organized Criticality (SOC). Significant progress has been made in  understanding SOC since the original work of Bak \emph{et al.} \cite{original}. The emergence of SOC is signaled by the births of anomalous avalanches, see \cite{16,17}  for recent work along these lines.
 There exists a new approach to SOC emphasizing temporal rather than intensity anomalous distributions \cite{lipiello,korosh}.  The authors of Ref. \cite{korosh} defined their approach to self-organization as \emph{Self-Organized Temporal Criticality} (SOTC). According to SOTC the crucial events defined earlier with the help of Eqs. (\ref{crucial1}) and  (\ref{crucial2}), namely the events that the authors of  \cite{allegrini}  were able to find in heartbeats, occur on an intermediate time scale, after an initial transient regime to the condition of intermediate asymptotics. The IPL nature of crucial events is tempered by an exponential relaxation in the long-time limit. This interpretation allows us to facilitate our approach to the connection between the
 diagnostic techniques of Ref. \cite{stanley} and  of Ref. \cite{allegrini}. In fact, the three time regimes
 of SOTC are a form of variability that we subsequently connect to the physiological variability
 that led the authors of Ref. \cite{stanley} to their diagnostic insight. 
 
 The search for crucial events is made difficult by the fact that crucial events are often imbedded in clouds of irrelevant events. The authors of Ref. \cite{allegrini} used a technique of statistical analysis, called Diffusion  Entropy Analysis (DEA) \cite{dea2}, to detect the anomalous scaling index $\delta$, which  these crucial events 
would generate were they not imbedded in a cloud of non-crucial events, namely, when they  are visible. According to the statistical analysis of Ref. \cite{allegrini} the distinction between healthy and pathologic subjects is established by noticing that the heartbeat dynamics of pathologic subjects host a critically large number 
of Poisson events.  An important result of this paper is the
observation that the Poisson events have the effect of reducing HRV. The largest  HRV is realized
in the ideal case of cardiac dynamics uniquely determined by the SOTC process, with its complete time evolution including the transient regime, intermediate asymptotics with its crucial events of Eqs. (\ref{crucial1}) and  (\ref{crucial2}), and the final tempered asymptotic  regime.

 Section \ref{intermediateasymptotics} affords intuitive arguments on the importance of intermediate asymptotics for the analysis of heartbeats illustrated in this paper.
 Section \ref{diffusionentropy} shows why DEA  works without being limited to the Gaussian condition.    In Section \ref{diffusionentropyandcrucialevents}, we show that the use of DEA adopted in
 earlier work \cite{allegrini} corresponds to
the observation of the intermediate asymptotic region. 
Section \ref{analysisofdata} reviews the procedure adopted in Ref. \cite{allegrini} to process the heartbeat data for the purpose of revealing, with the help of surrogate data,  to what  extent this is a genuine way of disclosing the contribution of Poisson events to the reduction of HRV.  Section \ref{joint} illustrates the joint use of DEA and detection of Poisson events. Finally, Section \ref{concludingremarks} is devoted to concluding remarks. 

\section{intermediate asymptotics} \label{intermediateasymptotics}
 In his book on intermediate asymptotics \cite{intermediate} Barenblatt adopts a visual art metaphor to illustrate the concept of intermediate asymptotics: ``... We have to look at paintings at a distance great enough not to see the brush-strokes, but at the same time small enough to enjoy not only the painting as a whole but also its important details: think of van Gogh's work, for example. ...".  Goldenfeld \cite{goldenfeld} illustrates the renormalization group rules that we have to adopt to eliminate the divergences created by the perturbation approach. This illustration is based on the assumption that  the physical condition of intermediate asymptotics is  a form of perennial transition to equilibrium.   
 
 There is a wide conviction that this is a simplifying but useful idealization of reality. A remarkable example is afforded by the work of Mantegna and Stanley \cite{mantegna}.  These authors noticed that although a finite size-induced truncation is an unavoidable consequence of the dynamics of real physical processes, the time duration of the transition to the Gaussian statistics prescribed by the central limit theorem may become extremely extended, in line with the idealized condition of perennial intermediate asymptotics of Goldenfeld.  However, for practical purposes a complex system can also be observed in  so large a time scale as to see dynamical effects that for simplicity may be interpreted as forms of ordinary fluctuation-dissipation processes.  Important work has been done to obtain analytical results for both short- and long-time regimes, see for instance \cite{koponen}, which triggered  significant interest in the appropriate mathematical formalism of transient anomalous diffusion \cite{randomwalk}, including exponential form of tempering \cite{tempering,majumdar}. It is convenient to notice that tempering may be an effect of representing real physical processes by means of finite length time series, an unavoidable consequence of observation. 
 We believe \cite{korosh} that tempering is a genuine property of the process of self-organization itself, since it emerges from the interaction of a finite number of units and that the heartbeat process belongs to this class of self-organizing processes, thereby involving tempering.

\section{Diffusion entropy} \label{diffusionentropy}
The DEA makes it possible to evaluate the correct scaling of a diffusion process, regardless of whether the Gauss condition applies or not \cite{scafettascaling}. The scaling $\delta$ of a diffusing variable $x$ is defined 
by
\begin{equation} \label{shannon}
p(x,t) = \frac{1}{t^{\delta}} F\left(\frac{x}{t^{\delta}}\right),
\end{equation}
where $p(x,t)$ is the PDF of the variable $x$  at time $t$ and $F(y)$ is a function that for crucial events does not have the ordinary Gaussian form.  DEA measures the  Shannon entropy of the diffusion process:
\begin{equation} \label{S(t)}
S(t) = - \int_{-\infty}^{+\infty} dx p(x,t) \ln \left[p(x,t)\right].
\end{equation}
By substituting Eq. (\ref{shannon}) into Eq. (\ref{S(t)}), after some algebra and replacing  the integration variable $x$ with the integration variable $y = x/t^{\delta}$, 
we obtain \cite{scafettascaling}
\begin{equation} \label{linear}
S(t) = A + \delta\ln (t),
\end{equation}
where the constant reference entropy is
\begin{equation}
A \equiv - \int_{-\infty}^{+\infty} dy F(y) \ln \left[F(y)\right].
\end{equation}

Eq. (\ref{linear})  shows that the entropy $S(t)$ increases linearly with $\ln(t)$  and the slope  of the resulting straight line is the scaling coefficient $\delta$.  The numerical search for the scaling coefficient   is done with this property in mind. Changing the unit adopted to measure time changes the value of $t$, but does not affect the scaling parameter $\delta$ \cite{scafettascaling}. DFA is based on evaluating scaling through the second moment of $p(x,t)$ and this has the effect of providing misleading information on $\delta$ when $p(x,t)$ has an IPL tail so slow as to generate divergence.  For this reason, Yazawa in his recent work on the effects of emotions on HRV adopted a modified version of DFA \cite{yazawa}. However, the MFDFA used herein is based on the adoption of fractional moments  $\left<|x|^q\right>$,  thereby bypassing the problems created by slow diffusion IPL tails with a conveniently small value of $q$.

\section{DEA as a technique to reveal crucial events}\label{diffusionentropyandcrucialevents}
The DEA method \cite{dea2} was originally introduced to properly analyze time series
assumed to be driven by crucial renewal events. 
It is important to stress that the renewal events hypothesized  \cite{allegrini} for the analysis of heartbeats are the subject of an extended literature focusing on the phenomenon of renewal aging \cite{burov}.  For a friendly illustration of the main results of this paper, we remind the readers about an algorithm used to generate the 
non-Poisson renewal events. It  is given by \cite{west,note}
\begin{equation} \label{step1}
\tau = T \left(\frac{1}{y^{\frac{1}{\mu-1}}} -1 \right), 
\end{equation}
where $y$ is a real number selected with uniform probability from the interval $(0,1)$.  The times $\tau$ generated by this algorithm are totally uncorrelated and obey the waiting time PDF
\begin{equation} \label{step2}
\psi(\tau) = (\mu -1) \frac{T^{\mu-1}}{(\tau + T)^{\mu}}.
\end{equation}
Note that to be as close as possible to the tempering prescriptions of 
SOTC \cite{korosh}, 
we should adopt a survival probability $\Psi(t)$ with the structure
\begin{equation}
\Psi(t) = \left(\frac{T}{t+ T}\right)^{\mu-1} \exp(-\Delta t),
\end{equation}
with the transient regime to intermediate asymptotics being 
determined by the parameter $T$ and defined by the time region $0 < t < T$.  The  time region of intermediate asymptotics corresponds to 
$T < t < \frac{1}{\Delta}$ and the tempered region is given by $t > \frac{1}{\Delta}$. For simplicity's sake the surrogate sequences hereby used are established using 
Eq. (\ref{step1}), which would correspond to $\Delta\to 0$, the tempered action being exerted by the finite size of the time series, $L$. We make the assumption that $\Delta \propto 1/L$.

In this paper, following the results of earlier work  \cite{allegrini}, we limit our analysis to the IPL index range:

\begin{equation}
2 < \mu < 3 .
\end{equation}
It is important to stress that the Poisson events correspond to $\mu = \infty$, but events drawn from $\mu = 5$ are already far enough from the crucial condition
as to be used safely as examples of non-crucial events.  The algorithm of Eq. (\ref{step1}) can be used to
explain in an intuitive way the different nature of the randomness
of $\mu < 3$ as compared to that of $\mu \gg 3$. The time interval
between two consecutive choices of the random number $y(0)$ has the mean value
\begin{equation}
\left<\tau\right> = \frac{T}{(\mu -2)},
\end{equation}
as can be easily established using the waiting time PDF $\psi(t)$ of Eq. (\ref{step2}) to perform the average. If $\left<\tau \right> < \Delta t$, where $\Delta t$ is the integration time step, we observe a process that is totally random. In the limiting case of $\mu < 2$, $\left<\tau \right> \gg \Delta t$, since in this case $<\tau> $ is divergent;  the randomness is sporadic. In the region $2 < \mu < 3$ randomness is not as sporadic as for $\mu < 2$. However, $<\tau^2> $ is divergent 
and as a consequence randomness remains distinctly intermittent. We make the assumption that the sporadic randomness of crucial events is good for the healthy function of cardiac dynamics and that an excess of randomness is risky. 

To discuss the joint action of frequent and sporadic randomness let us  create suitable surrogate time series, namely an appropriate sequence of times 
$\tau_1, \tau_2, ....\tau_i, \tau_{i+1}, ....$. This sequence is generated by a repeated random selection of $y$ of Eq. (\ref{step1}) so as to create either a sequence of crucial events, with $\mu < 3$,
or a sequence of non-crucial events, with $\mu > 3$. More precisely, in the applications of this paper we adopt $3 > \mu > 2$ for crucial events and $\mu = 5$ for non-crucial events. 

Each of these two  time sequences has to be turned into a corresponding suitable fluctuation $\xi(t)$. To do that we adopt   the Asymmetric Jump Model (AJM) \cite{dea2}. The reason for this choice, illustrated in detail in Ref. \cite{dea2},
is that this random walking rule makes it possible for DEA to reveal the correct scaling established by the generalized central limit theorem (GCLT) \cite{feller} in the whole crucial event region $\mu < 3$, including the region $\mu < 2$. This walking rule is established by setting $\xi = 0$ when there are no events, 
and $\xi = 1$ when either a crucial or Poisson event occurs.

Thus we create two time series, one corresponding to $\mu < 3$
and one corresponding to $\mu > 3$. The surrogate time series used here for the statistical analysis  corresponds to the superposition of both time series,
\begin{equation} \label{surrogate}
\xi(t) =(1-\epsilon) \xi_{\mu > 3}(t) + \epsilon \xi_{\mu < 3}(t). 
\end{equation}
The parameter $\epsilon < 1$ is the probability that the observed heartbeat  signal, detected according to the prescription of the next section is generated by a genuine SOTC process.  In Section \ref{analysisofdata} we  explain how to derive $\epsilon$ from the analysis of real heartbeat data. 

In the case where SOTC events are visible, namely $\epsilon = 1$, the method of DEA leads to the detection of the proper scaling
\begin{equation} \label{crucialscaling}
\delta = \frac{1}{\mu -1}
\end{equation}
after an initial transient corresponding to the micro-time regime, where the complexity of the process is not yet perceived. 
Notice that the transition from the L\'{e}vy to the Gauss regime occurs at $\mu = 3$. However, as stated earlier, the surrogate time series of this paper rest on $\mu = 5$, namely a condition well imbedded in the Gaussian basin of attraction. 

\begin{figure}[ptb]
\begin{center}
\includegraphics[width=1\linewidth]{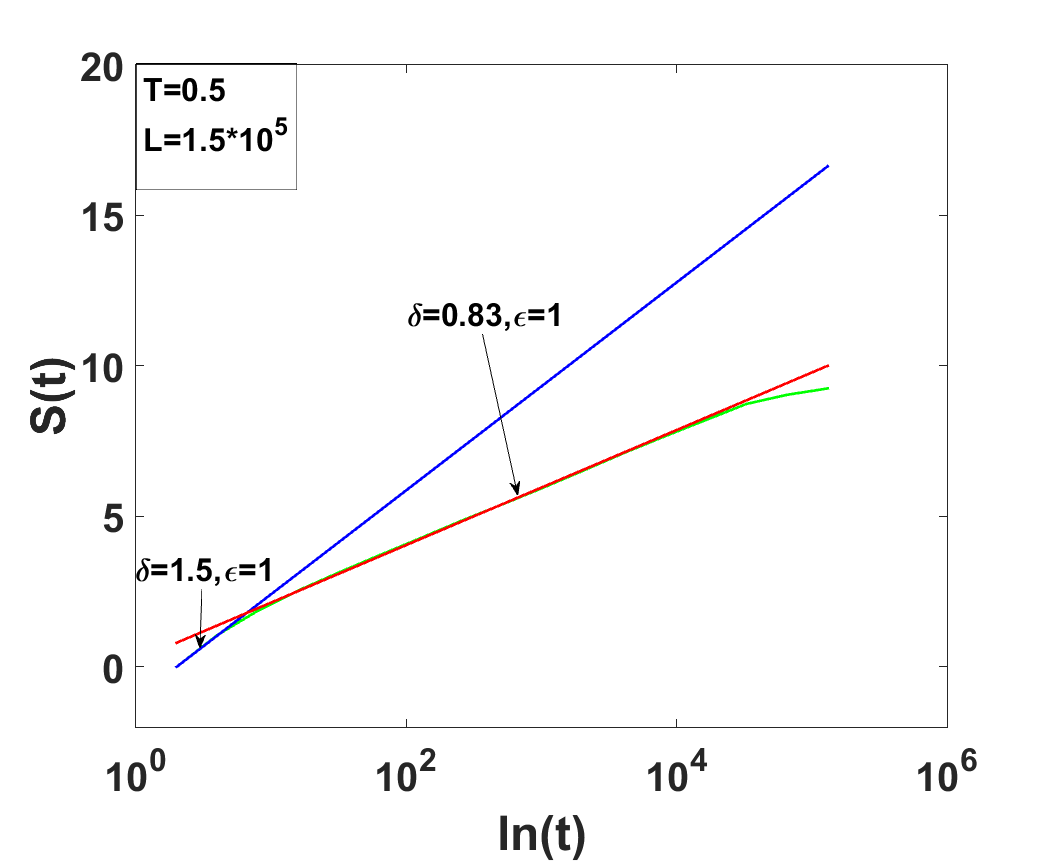}
\end{center}
\caption{Entropy of the time series versus the logarithm of time from the micro-time to the asymptotic time scale with $\epsilon = 1$.}
\label{giacomo1}
\end{figure}

\begin{figure}[ptb]
\begin{center}
\includegraphics[width=1\linewidth]{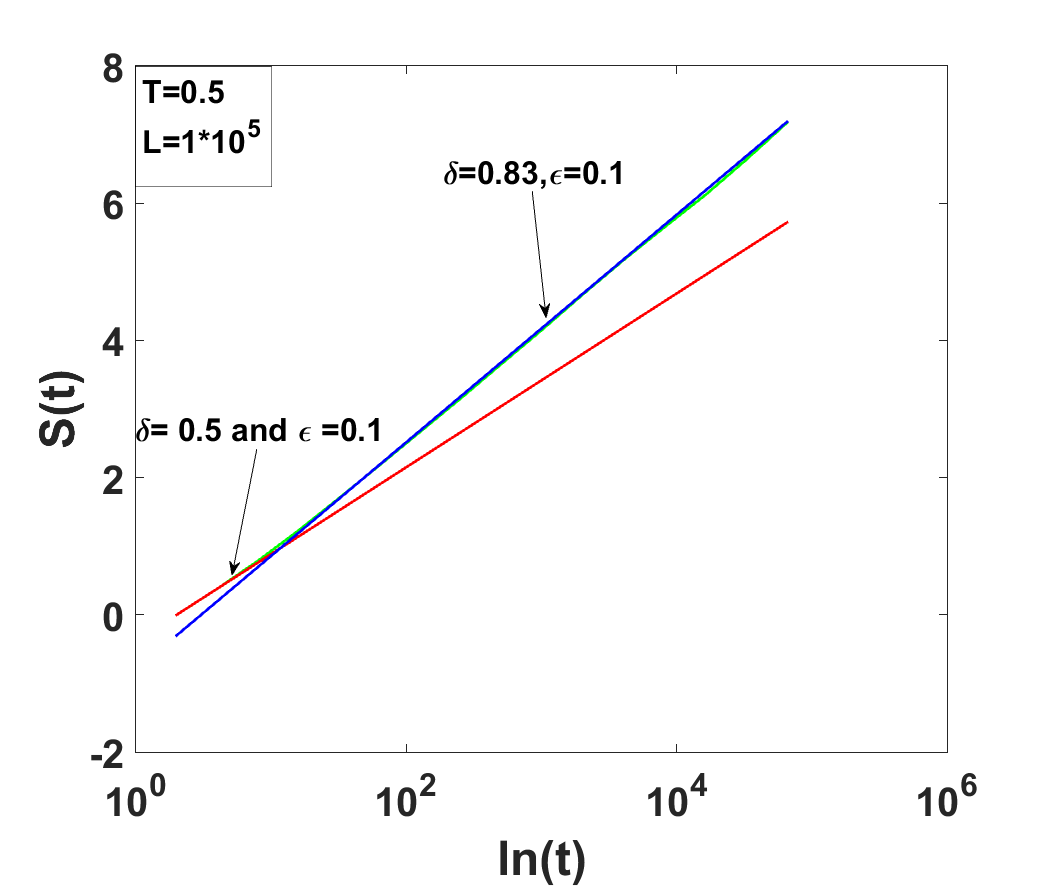}
\end{center}
\caption{Entropy of the time series versus the logarithm of time from the micro-time Gaussian basin of attraction to the asymptotic time scale with $\epsilon = 0.1$.}
\label{giacomo2}
\end{figure}

\begin{figure}[ptb]
\begin{center}
\includegraphics[width=1\linewidth]{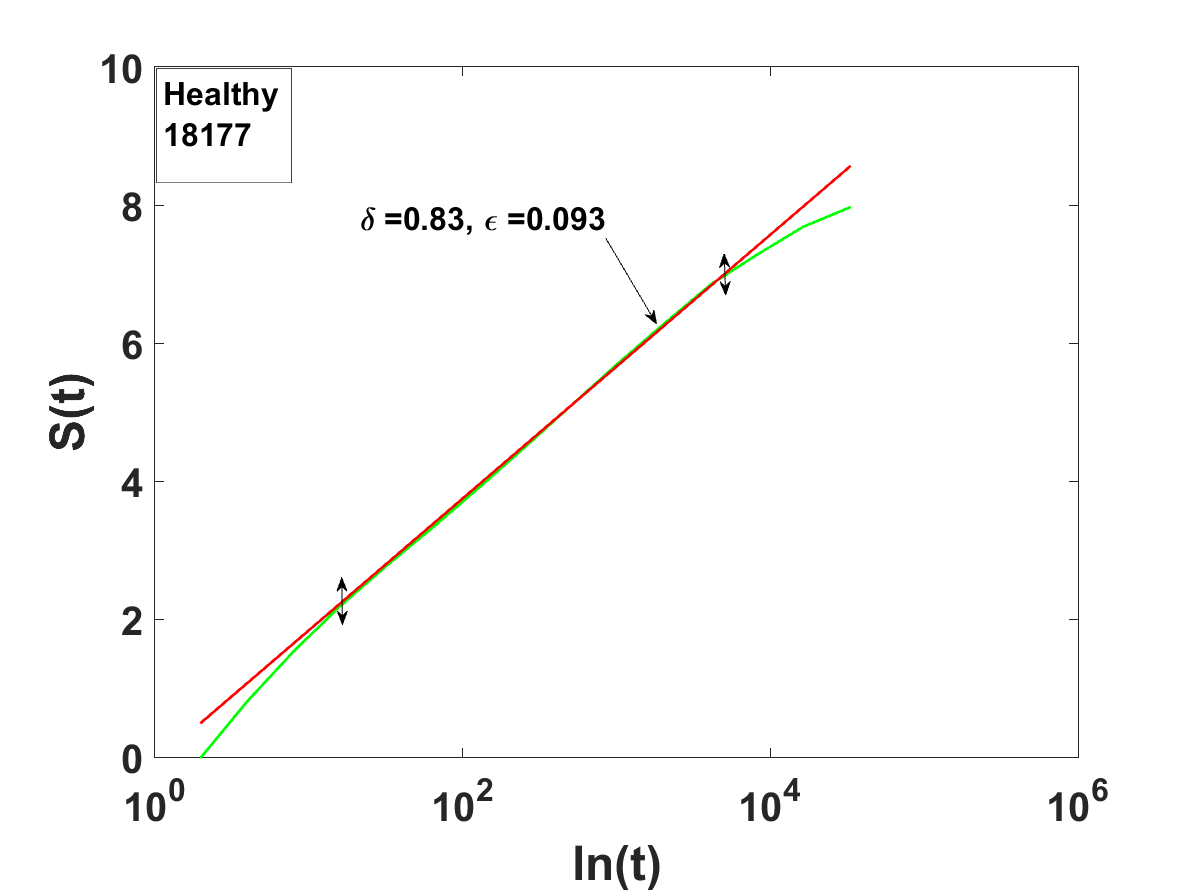}
\end{center}
\caption{DEA detects the scaling of invisible crucial events in the intermediate asymptotic time. The scaling $\delta$ is the slope of the straight line between the two vertical arrows. }
\label{giacomo3}
\end{figure}

Figs. \ref{giacomo1},  \ref{giacomo2} and \ref{giacomo3} show DEA in action,  through the linear-log representation, which is used, according to Section \ref{diffusionentropy}, to detect the scaling $\delta$, the slope of the linear portion of $S(t)$ in this representation. 

Fig. \ref{giacomo1} illustrates the case where $\epsilon = 1$, namely the condition where the crucial events are fully visible, with $\mu = 2.2$. The corresponding crucial scaling should be $\delta = 0.83$. However, in  the short time regime 
the scaling  has the larger value $\delta = 1.5$ and the scaling $\delta = 0.83$ of crucial events appears in the intermediate time regime.  For this reason, the proper scaling,
as shown in this figure, is optimal in the intermediate time regime. Actually, we see that in the region around $t \propto 10^5$ a tempering deviation from the the crucial scaling of Eq. (\ref{crucialscaling}) occurs. This is not the tempering of the 
SOTC defined in Ref. \cite{korosh}. The theoretical study of that physical tempering of the process is outside the scope of the present paper, but we make the reasonable assumption that
heartbeat dynamics fit it as a consequence of being itself a process of self-organization.

Fig. \ref{giacomo2} illustrates the more important case where
the crucial events are hidden by a cloud of noncrucial events.
In this case, too, according to earlier analysis \cite{allegrini},
the correct scaling generated by the crucial events appears in the intermediate time regime. However, in this case the reason for the initial transient is quite different from the SOTC initial transient. In this case the initial short-time regime characterized by the conventional scaling 
$\delta = 0.5$, corresponds to the scaling of Poisson events.  In the long-time regime, when the
SOTC intermediate asymptotic emerges,  the faster scaling of the crucial events with $\mu < 2$ leads them to crossover to ordinary diffusion. 
The overlap of the Poisson-induced transient regime 
and transient SOTC make the derivative of the diffusion entropy non-monotonic. For simplicity's sake we do not show this complicated behavior, instead we focus on the complexity of the intermediate asymptotics. 
Notice that, although the extended transient to the
intermediate asymptotic regime induced by a large percentage of Poisson events can be confused with the transient SOTC regime, the corresponding physical effects are the opposite of one another. The SOTC transient 
generates a broad multi-fractal spectrum, while the long transient induced by a large percentage of Poisson events has the effect of making the multi-fractal spectrum narrower.

To complete the discussion of this section we make some comments concerning Fig. \ref{giacomo3}. In Section \ref{analysisofdata} we  explain how to derive this figure from real data on heartbeats. Here we limit our observation to the scaling $\delta$, representing the indicator of the occurrence of crucial events. The IPL index is evaluated by monitoring the intermediate asymptotics region, the short- and long-time limit of which are denoted by vertical arrows. In this case, the deviation from Eq. (\ref{crucialscaling}) of the tempering region
is probably due the properties of heartbeats, rather than to the finite size $L$ of the sequence under study.

In summary, it is important to reiterate that on the basis of recent advances made concerning SOTC \cite{korosh}, the time series generated by complex processes are characterized by three regimes: the short-time regime,
where the true complexity of the process is not yet perceived; an intermediate time regime driven by the crucial events; and a long-time regime where the process can be mistaken for an ordinary statistical process. It is on the contrary a tempering effect generated by self-organization.

\section{How to process experimental data to reveal the existence of crucial events} \label{analysisofdata}

Following  \cite{8} and \cite{allegrini}, we use the  ECG records of the MIT-BIH Normal Sinus Rhythm Database and of the BIDMC Congestive Heart Failure Database, for  healthy and  congestive heart failure  patients, respectively.

\begin{figure}[ptb]
\begin{center}
\includegraphics[width=1\linewidth]{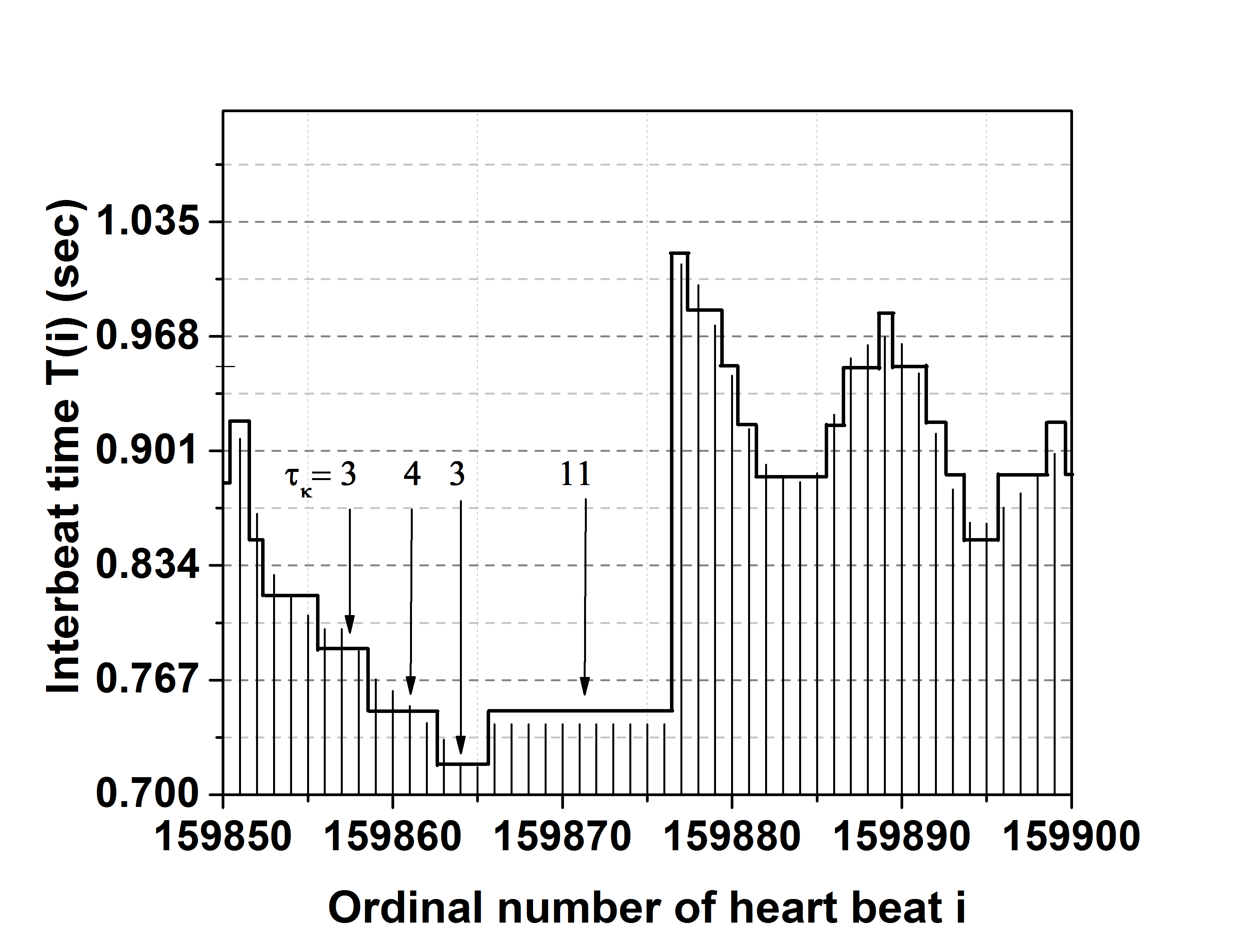}
\end{center}
\caption{Rule adopted to define events. An event is defined as the experimental curve, thick black line, crossing the border between two consecutive strips. The size of the strips is $\Delta T = 1/30 $ sec.  }
\label{stripes}
\end{figure}

\begin{figure}[ptb]
\begin{center}
\includegraphics[width=1\linewidth]{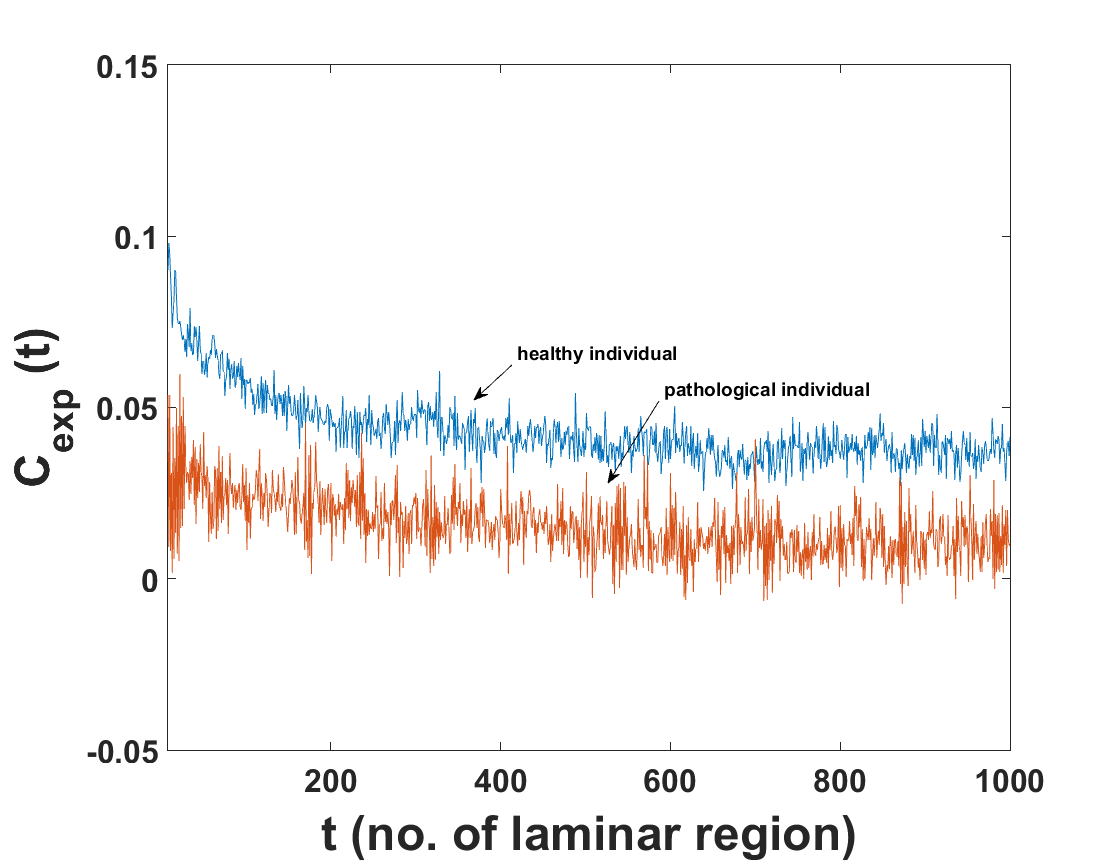}
\end{center}
\caption{Correlation function $C(t)$ for two typical patients, one healthy and one pathological.}
\label{experimentalCORRELATION}
\end{figure}

The main problems encountered in proving that SOTC is the process driving the phenomenon under study has to do with the detection of the crucial events, namely, events with a waiting time PDF yielding a diverging second moment.
Fig. \ref{stripes} shows the approach we adopt, identical to that used in  Ref. \cite{allegrini}. The experimental signal is obtained by assigning to each beat a value corresponding to the time interval between one and the next. We divide the inter-beat time axis into small strips of  size $\Delta T$. We define the occurrence of an event as  the experimental signal crossing from one strip to one of the two nearest neighbor strips. We see that the heartbeat trajectory may remain in a given strip for an extended time, suggesting the typical intermittent behavior that led to the
discovery of crucial events. However, the experimental signal crossing the border between two contiguous strips is not necessarily a crucial event. The crucial events are renewal and consequently the times $\tau_i$ should not be correlated. To assess the breakdown of the renewal condition we evaluate the time-average correlation function, where the time average is indicated by an overbar
\begin{equation}
C(t) = \frac{\sum\limits_{|i-j| = t}\overline{\left(\tau_i - \overline {\tau}\right)\left(\tau_j - \overline {\tau}\right)}}{\sum\limits_i\overline{\left(\tau_i - \overline {\tau}\right)^2}}.
\end{equation}
This correlation function is properly normalized, thereby yielding $C(0) = 1$, and in the case of genuine renewal events should fit the condition
$C(t) = 0$ for $t> 0$. 
Fig. \ref{experimentalCORRELATION} shows, on the contrary,  that the correlation function $C(t)$ makes an abrupt jump from $1$ to a very small, but non-vanishing value  of $\epsilon^2$,  suggesting that  the technique adopted to reveal events actually does not detect genuine renewal events.
Notice that for a proper definition of $\epsilon$ we define $\epsilon^2$ either as the value of $C(1)$, if $C(1) > 0$ or the mean value over the first one hundred events, if $C(1) < 0$.

\begin{figure}[ptb]
\begin{center}
\includegraphics[width=1\linewidth]{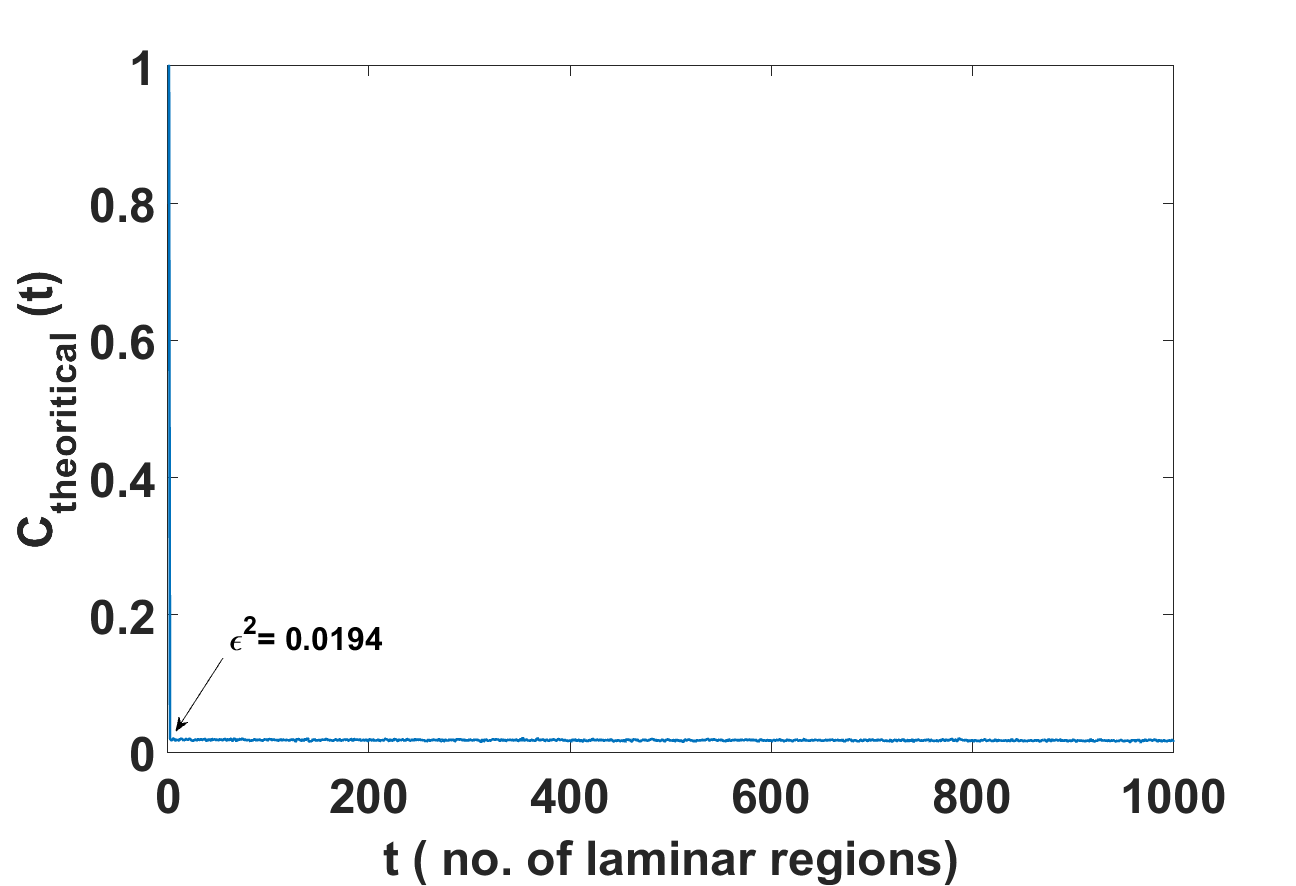}
\end{center}
\caption{Correlation function C(t) for the surrogate data in the case of strong randomness.}
\label{strongRANDOMNESS}
\end{figure}

\begin{figure}[ptb]
\begin{center}
\includegraphics[width=1\linewidth]{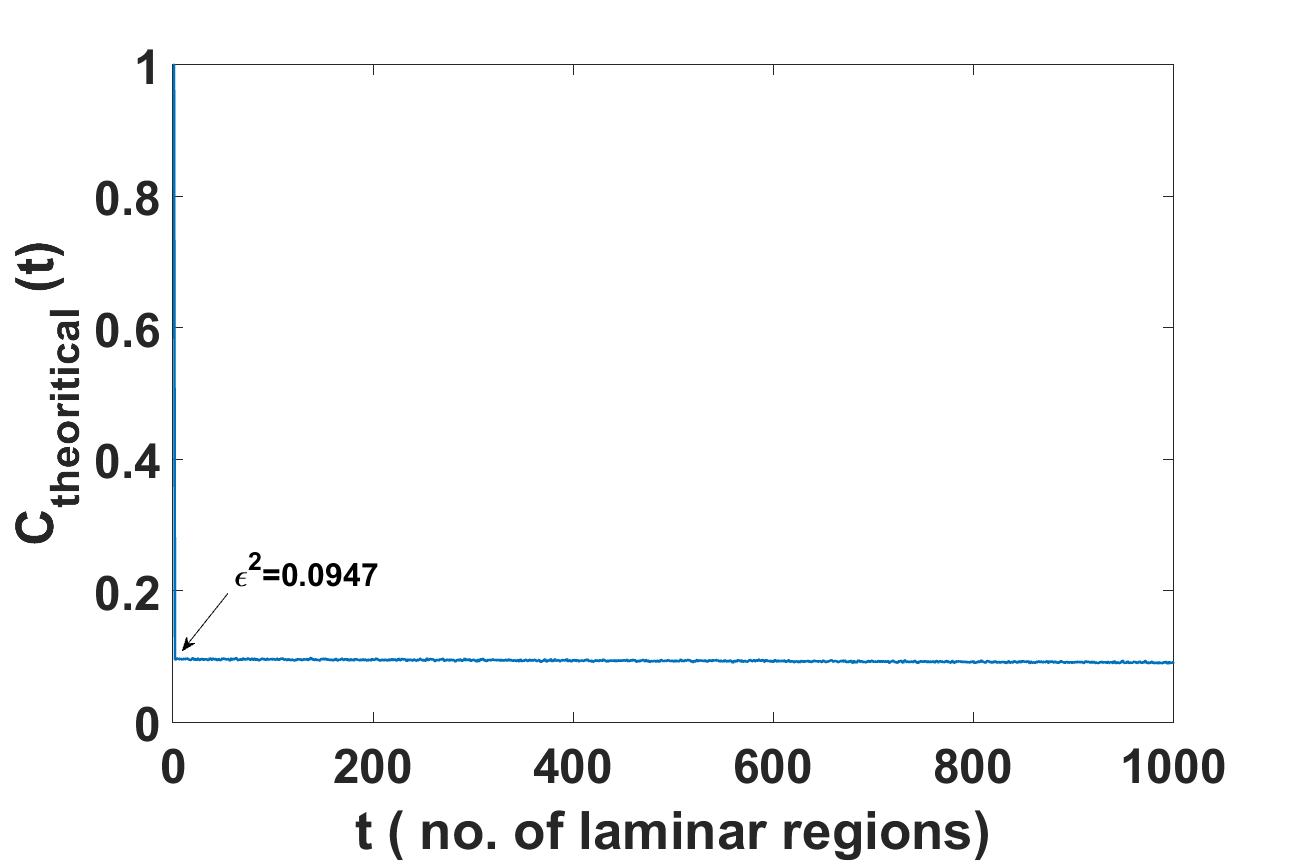}
\end{center}
\caption{Correlation function C(t) for the surrogate data in the case of weak randomness.}
\label{weakRANDOMNESS}
\end{figure}

To understand the meaning of $\epsilon^2$ we interrogate the surrogate sequences defined by Eq. (\ref{surrogate}).
With the help of Fig. \ref{strongRANDOMNESS} and Fig.\ref{weakRANDOMNESS} we establish, in line with  earlier work \cite{allegrini},  that  the intensity $\epsilon^2$ is the square of the probability that an event is a crucial event. This explains why we adopt the symbol 
$\epsilon^2$ to denote the value of the correlation C(t) immediately after the abrupt jump down from $C(0) = 1$. 
Fig. \ref{strongRANDOMNESS} shows a theoretical correlation function using a surrogate sequence in action for strong randomness.
Fig. \ref{weakRANDOMNESS} shows a theoretical correlation function using a surrogate sequence in action for weak randomness.

\section{Joint use of DEA and C($t$)}\label{joint}

In this Section, we recover the central result of Ref. \cite{allegrini}, which was based on the joint use of DEA and the correlation function  $C(t)$. 
For each subject we define both $\delta$ and $\epsilon^2$. 

\begin{figure}[ptb]
\begin{center}
\includegraphics[width=1\linewidth]{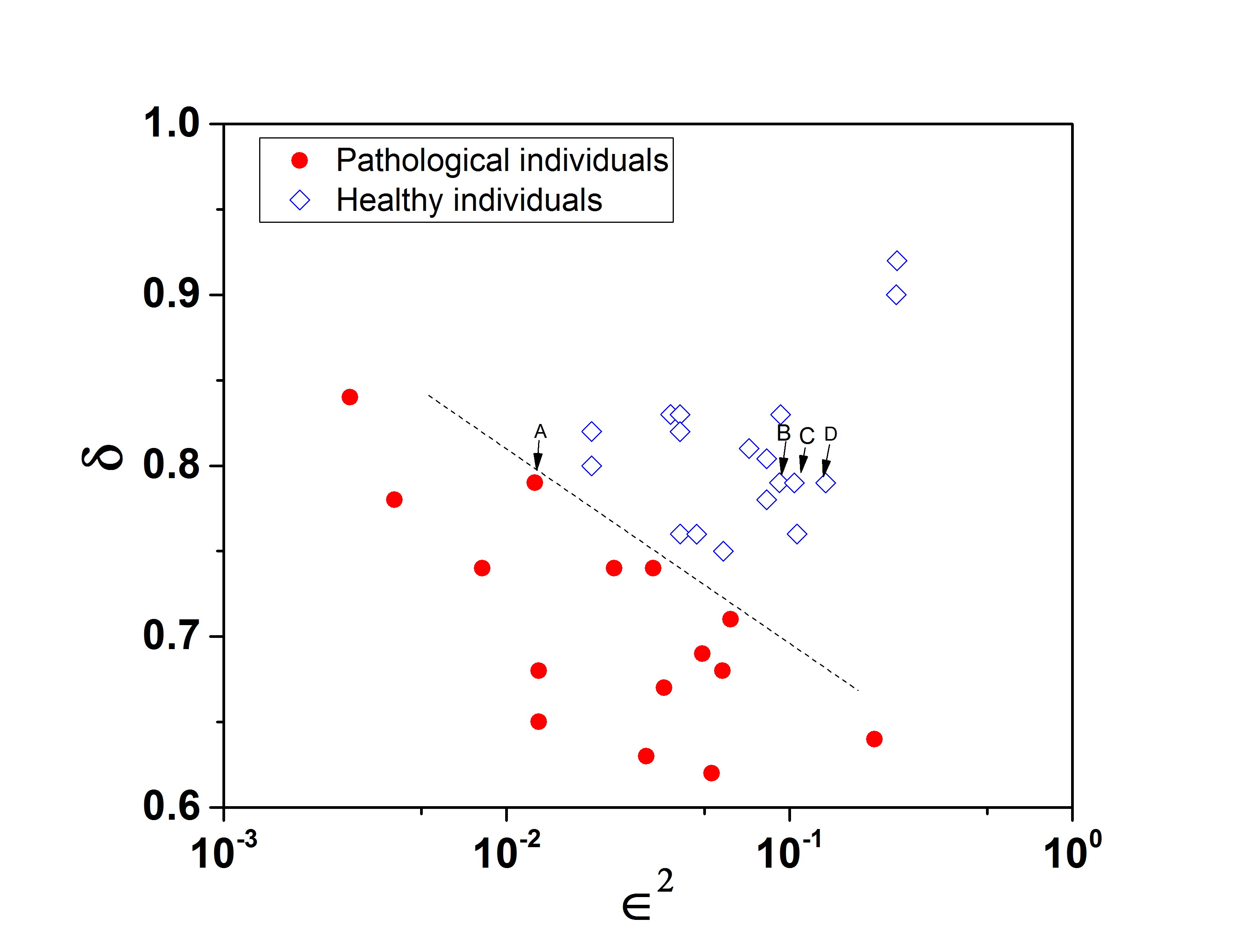}
\end{center}
\caption{Distinguishing subjects with healthy from those with pathological HRV.}
\label{recoveringALLEGRINI}
\end{figure}

In fact, Fig. \ref{recoveringALLEGRINI} is virtually identical to the central result found by the authors of Ref. \cite{allegrini}, which establishes a criterion to distinguish patients with pathological from those with healthy HRV. We notice that the ideally healthy condition would correspond to $\epsilon = 1$ and $\delta = 1$. This means
that the crucial events should not host any Poisson event and should have $\mu = 2$, which is the border between the region of perennial aging, $\mu < 2$, and
the region where the rate of randomness production becomes constant in the long-time limit, $\mu > 2$ \cite{west}. 
The patient HRVs move toward the pathological condition as their scaling becomes closer to the scaling of ordinary diffusion $\delta = 0.5$, namely closer to the border between the region of crucial events, $\mu < 3$, and the Gaussian basin of attraction, $\mu > 3$ . 

Note that the work of Ref. \cite{brain} established that the brain, generating ideal $1/f$-noise, is located at the border between the region of perennial aging and the region
of crucial events hosted by heartbeats,  according to the analysis of this paper
and  earlier work \cite{allegrini}.  

The research work done in the new field of network medicine \cite{ivanov}  focuses on the interaction between the different organs of human body,  the brain and heart being  a special case of this intercommunication \cite{brainandheart}.  According to the theory of \emph{complexity matching} \cite{west}, based on the assumption that the synchronization of complex networks is facilitated by the networks sharing the same complexity, $\mu = 2$, in the case of brain-heart communication, we make the plausible conjecture that the right-top corner of Fig. \ref{recoveringALLEGRINI} corresponds to a convenient condition for brain-heart communication in the ideal case of healthy patients. However, the current literature on complexity matching emphasizes the communication between the two complex networks through their multi-fractal spectra \cite{deligniers}.  Therefore, establishing a connection  between crucial events and multi-fractal spectra 
is a goal of this paper. The most important property
of Fig. \ref{recoveringALLEGRINI} is to contribute to the realization of that goal by establishing a connection  between \cite{allegrini} and \cite{8,stanley}.

We focus our attention on the individuals labeled A, B, C and D in Fig. \ref{recoveringALLEGRINI}. These patients have the same $\delta$ and  according to the earlier analysis \cite{allegrini} the distinction between sick and healthy patients is due to the fact that 
the heartbeat of the sick patients is affected by excessive randomness.  According to the authors of Ref. \cite{8,stanley} the distinction is due to the fact that healthy patients have  broader multi-fractal spectra. 

The central result of the present paper is obtained by applying the MFDFA to the individuals A, B, C and D
for the  purpose of proving the connection between the diagnostic recipe of Ref. \cite{allegrini} and that of Refs. \cite{stanley,8}.

\begin{figure}[ptb]
\begin{center}
\includegraphics[width=1\linewidth]{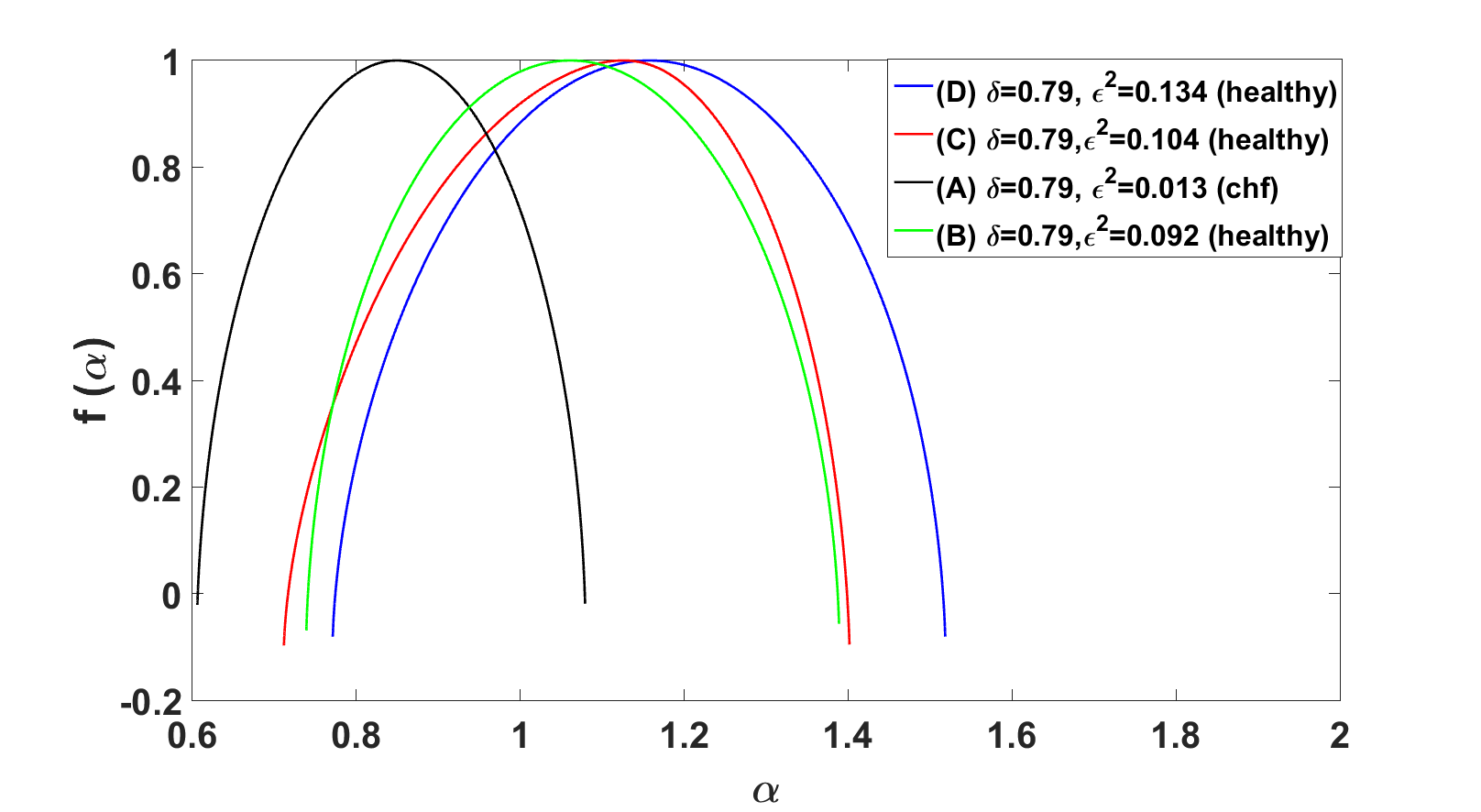}
\end{center}
\caption{Multi-fractal spectra of HRV as a function of $\epsilon$ keeping constant the crucial scaling $\delta = 0.79$.}
\label{centralRESULT}
\end{figure}
\begin{figure}[ptb]
\begin{center} \includegraphics[width=1\linewidth]{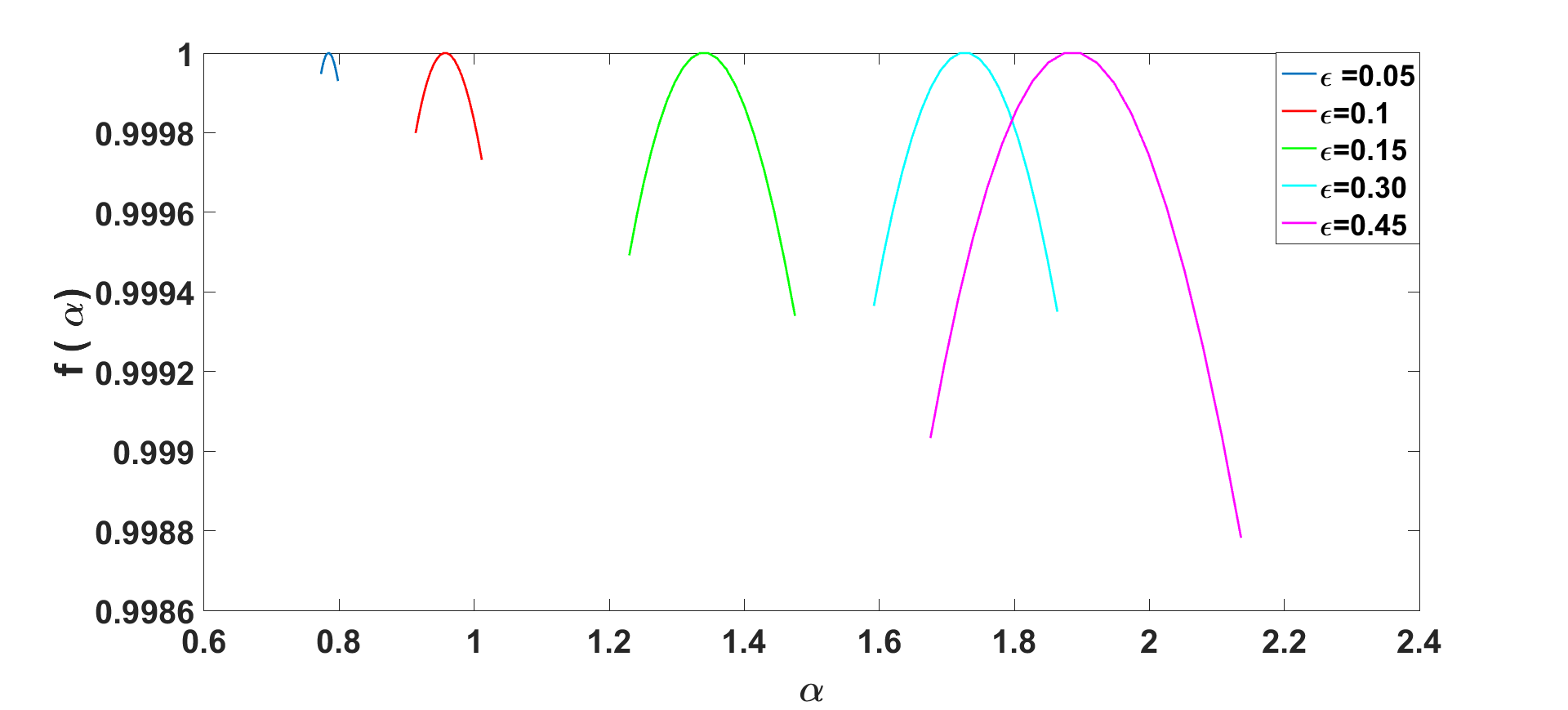}
\end{center}
\caption{Multi-fractal spectra of surrogate data as a function of $\epsilon$ keeping constant the crucial scaling $\delta = 0.83$. }
\label{centralCENTRAL}
\end{figure}

Fig. \ref{centralRESULT} fully confirms this connection. We see, in fact, that moving from the sick to the healthy patients has the effect of increasing the width of  the multi-fractal spectrum. 
Note that Fig. \ref{centralCENTRAL} provides additional confirmation of this connection through the use of surrogate sequences.

\section{CONCLUDING REMARKS} \label{concludingremarks}

\begin{figure}[ptb]
\begin{center} \includegraphics[width=1\linewidth]{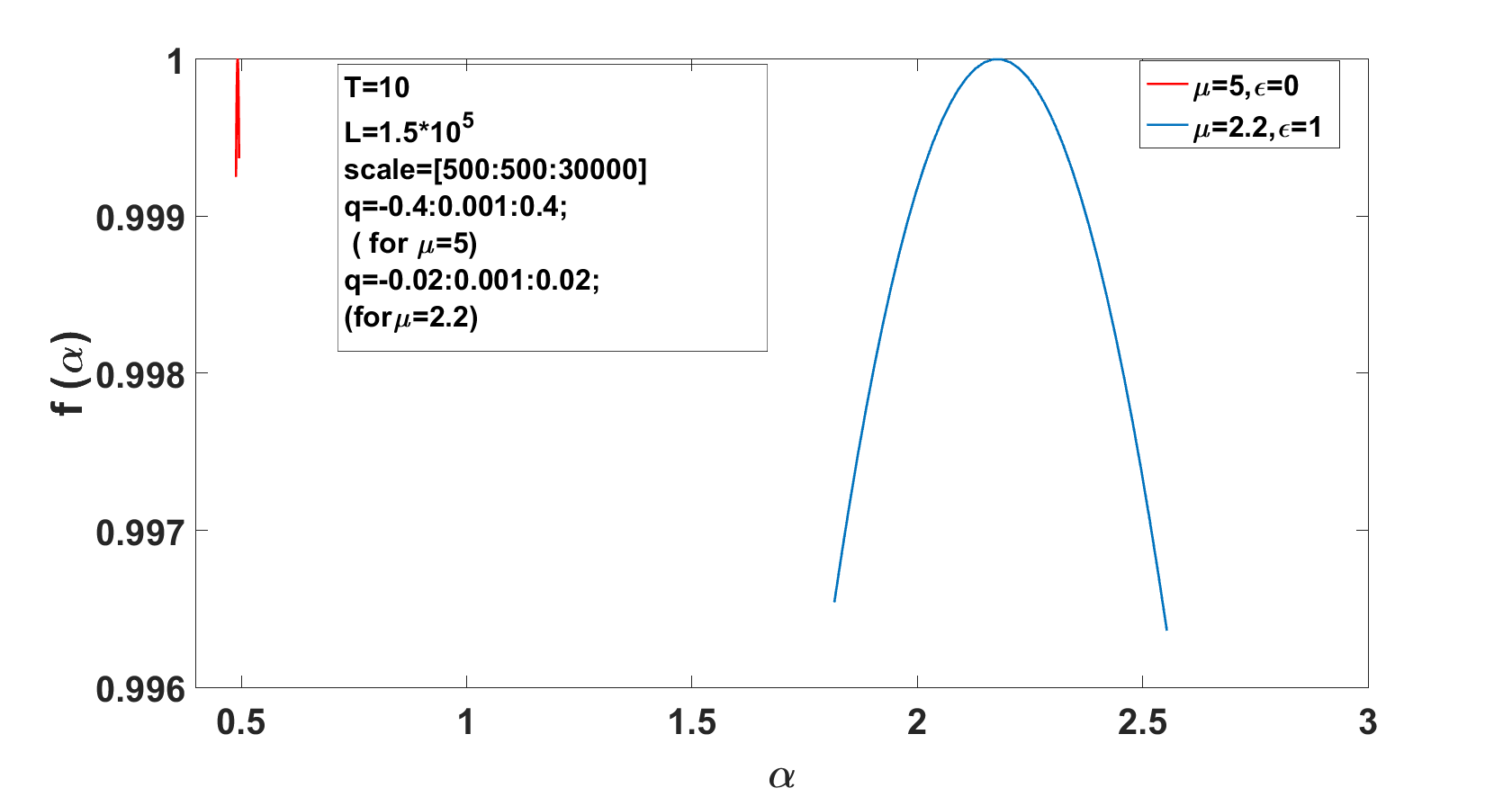}
\end{center}
\caption{Extreme cases of most narrow, $\epsilon = 0$, and most broad, $\epsilon = 1$, multi-fractal spectra.}
\label{final}
\end{figure}

The diagnostic method generated by the work of Ref. \cite{allegrini} yields some benefit compared to that of Ref \cite{stanley}. One of these benefits is that the distinction between healthy and pathologic patients is established through the two-dimensional representation of Fig. \ref{recoveringALLEGRINI} rather than the three-dimensional representation of Ref. \cite{stanley}. 
Another important result of this paper is its contribution to an improved vision of variability and multi-fractality.  
To  appreciate this significant improvement let us focus our attention on the results obtained by applying  the MFDFA to the surrogate  series in the limiting case
of a visible SOTC \cite{korosh} process, $\epsilon = 1$,  and of  
a Poisson process, $\epsilon = 0$. The result of this analysis is shown in Fig. \ref{final}. The narrowest multi-fractal spectrum is realized by setting $\epsilon = 0$.
We reiterate that, according to SOTC \cite{korosh}, crucial events are characterized by three distinct time regimes, a transient initial regime, the intermediate asymptotics time regime, and a final tempered time regime with exponential truncation. The transient 
time regime becomes more and more extended with decreasing $\epsilon$. However,  the extended
transient regime generated by a very small value of 
$\epsilon$ must not be confused with a wide transient regime
corresponding to the occurrence of a sufficient number of crucial events to realize the prescription $\delta = 1/(\mu-1)$ of the GCLT \cite{dea2,feller}.  The GCLT transient regime is
the micro-evolution towards the IPL regime predicted by SOTC \cite{korosh}. This transient regime, the intermediate asymptotic time regime and the final tempering time regime are the generators of the wide variability that the multi-fractal DFA efficiently detects.
The Poisson events generated by $\mu \gg 3$ generate an extended transient regime that has the opposite effect of yielding an extremely narrow spectrum around the ordinary scaling value 
$\alpha = 0.5$.

In conclusion, the results of the present paper establish a clear connection between the multi-fractal spectrum and SOTC fluctuations, thereby affording 
a promising tool to make further progress in the field of network medicine \cite{ivanov}, where broad multi-fractal spectra are transferred, according to \cite{west}, from one network to another via crucial events.

\emph{Acknowledgment} The authors thank Welch and ARO for financial support through Grant No. B-1577 and W911NF-
15-1-0245, respectively.


.


\begin{references}
\bibitem{note} Eq. (\ref{step1}) is Eq. (180) of \cite{west} with  $(\mu-1$ ), misprint, correctly replaced by $1/(\mu-1)$.
\end{references}
\end{document}